%% file: main.tex
\documentclass[twocolumn,10pt]{article}
\usepackage[letterpaper,left=1.95cm,right=1.95cm,top=2.5cm,bottom=3.16cm]{geometry} 
\input{Styles_Eje.sty}  
\graphicspath{ {images/} }

\usepackage{blindtext}
\usepackage{subfiles} 

{\title{\fontsize{16}{16}
\vspace*{-0.7cm}
\textbf{
Quantum computing overview: discrete vs. continuous variable models
} \\[0.2cm]
\fontsize{12}{12}
\textbf{superconducting vs. linear optics}\\[0.2cm]}
}

\author[1]{\fontsize{10pt}{10pt}\selectfont \textbf{Sophie Choe}}

\affil[1]{
\fontsize{8pt}{8pt}\selectfont Electrical and Computer Engineersing, Portland State University, Portland, OR}

\begin{document}
\input{Abstract}

\subfile{intro}

\subfile{qc}

\subfile{imp}

\subfile{QPU}

\subfile{conc}

\bibliographystyle{plain}

\subfile{ref}
\end{document}

%% file: Abstract.tex
\twocolumn
[
\begin{@twocolumnfalse}
\maketitle
\begin{abstract}
\vspace*{0.5cm}
\justify
\fontsize{10pt}{10pt}\selectfont
In this Near Intermediate-Scale Quantum era, there are two types of near-term quantum devices available on cloud: superconducting quantum processing units (QPUs) based on the discrete variable model and linear optics (photonics) QPUs based on the continuous variable (CV) model. Quantum computation in the discrete variable model is performed in a finite dimensional quantum state space and the CV model in an infinite dimensional space. In implementing quantum algorithms, the CV model offers more quantum gates that are not available in the discrete variable model. CV-based photonic quantum computers provide additional flexibility of controlling the length of the output vectors of quantum circuits, using different methods of measurement and the notion of cutoff dimension.

\keywordsEng{quantum computing, quantum machine learning, quantum neural networks, continuous variable quantum computing, photonic quantum computing, classical quantum hybrid model, quantum MNIST classification}
\end{abstract}
\vspace{0.5cm}

\hspace*{0.7cm}
\textit{Dated: April 2022} \\
\hspace*{0.7cm}
\vspace*{0.5cm}

\end{@twocolumnfalse}
]

%% file: intro.tex
\section{INTRODUCTION}

\justifying
The current landscape of quantum computing is defined as Near Intermediate-Scale Quantum or near-term quantum \cite{NISQ_18}, where functioning QPUs are available on cloud for end users. These QPUs are called near-term devices and they are characterized by small-scale and shallow circuits \cite{circuit_18} to minimize hardware errors, until they become fully fault-tolerant and universal. 

Nonetheless, the availability of working QPUs offers the opportunity to run and test quantum algorithms, which were only theoretical previously. Research on quantum algorithms specific for these devices has emerged as an active area of research, especially in quantum chemistry, Gaussian boson sampling, graph optimization, and quantum machine learning. Unlike the original assumption that quantum computers would replace classical computers, QPUs are emerging as task-specific special processing units, much like Graphical Processing Units.

There are two types of near-term quantum devices available for end users: superconducting quantum computers based on the discrete variable (qubit-based) model and photonic quantum computers based on the continuous variable model.

Superconducting computers use the particle-like property of nature and utilizes individual particles as information carriers such as electrons. The possible states a particle can be found in are up and down, identified with 0 and 1 states, hence the computational state in the qubit model is finite. In order to keep them in stable states that can be described quantum mechanically, we need to maintain the ambient temperature very low, hence the chips need to be inside a dilution refrigerator. Each particle implementing a quantum bit needs to be connected to the corresponding hardware implementing a classical bit with a wire and controlling the temperature of the wire as not to disturb the quantum states of qubits needs sophisticated engineering \cite{qc_19}.

Photonic computers, on the other hand, use bosonic modes (qumodes) as information carriers and they can be created and maintained at room temperature using linear optics instruments. Hence they can easily be incorporated into existing computing infrastructure. The computational state space under the CV model, on which photonic computers are based on, is infinite dimensional with a richer array of quantum gates than in the qubit model.

This paper provides an overview of quantum computing, the difference between the discrete variable model and the CV model, and available QPUs based on these models.

%% file: qc.tex
\section{QUANTUM COMPUTING}
Quantum computing is a paradigm of computing using physical systems that are governed by quantum mechanics. Hence the evolution of state changes, representing information processing, can take advantage of quantum mechanical effects not present in classical mechanics \cite{qi_10}. Most importantly, the mathematical formalism describing quantum mechanical systems is based on a higher dimensional quantum state space. The computational space in quantum computing is of higher degree than in classical computing. Therefore, some computations impossible in classical computational spaces can be actualized in quantum or some classical algorithms can be processed much faster.

In the gate model quantum computing, where information processing is carried out by quantum logic gates, the quantum states of individual information carriers are used as units of information. This is implemented by some physical system whose state can be expressed quantum mechanically, in the superconducting model by individual particles and in the linear optics model by bosonic modes \cite{qc_19}.

The transition from classical to discrete variable quantum can be viewed as an extension of computational space from the two element space $\{0, 1\}$ to a higher dimensional projective Hilbert space $\mathbb{C}P^1 \equiv \mathbb{C}^2$ modded out by some projective identification. The transition from discrete variable quantum to continuous variable quantum can be viewed as an extension of computational space from finite dimensional $\mathbb{C}P^1$ to infinite dimensional $\mathbb{C}P^{\infty}$. The corresponding the computational basis goes from $\{\ket{0}, \ket{1}\}$ to $\{\ket{0},\ket{1}, \ldots, \ket{n}, \ldots \}$ \cite{qi_cv_03}. In addition to an infinite dimensional computational space of the CV model, the computational basis states $\ket{0}, \ket{1}, \ldots$ are quasi-Gaussian distributions as opposed to single points. Under the mathematical formalism of quantum mechanics, any possible state of a quantum mechanical system is represented by a unit vector in a projective Hilbert space, $\mathbb{C}P^1$ in the qubit model and $\mathbb{C}P^{\infty}$ in the CV model. The space of complex numbers $\mathbb{C} = \mathbb{R} \times i\mathbb{R}$ can be viewed as $\mathbb{R}^2 = \mathbb{R} \times \mathbb{R}$, hence embedding of classical information processing in quantum automatically expands the computational space. 

Parallelism because of superposition.\\
Any quantum state in a qubit computational space is in superposition of the computational basis states $\ket{0}=\begin{bmatrix} 1 \\ 0 \end{bmatrix}$ and $\ket{1}=\begin{bmatrix} 0 \\ 1 \end{bmatrix}$ simultaneously allowing for parallel processing. In CV quantum computing, the computational basis is of an infinite size: $\{\ket{0}, \ket{1}, \ldots, \ket{n}, \ldots \}$, increasing the degree of parallelism substantially.

A Hilbert space is a vector space equipped with the inner product operation. With the inner product property of a Hilbert space, we can express the distance and the angle between two vectors, using Bra-ket notation, also known as Dirac notation. The notation ket $\ket{x}$ is used to denote a column vector with complex entries and the notation bra $bra{x}$ to denote its complex conjugate transpose, which becomes a row vector. The distance between two vectors is defined as the length of the difference vector of the two.
\begin{equation*}
\begin{aligned}
dist(\ket{x}, \ket{y}) &= \Vert \ket{x-y} \Vert \\
&= \sqrt{\Vert \ket{x-y} \Vert ^2}\\
&= \sqrt{\sum_{k=0}^{n-1}\Vert x_{k} - y_{k}\Vert ^2}\\
&= \sqrt{\sum_{k=0}^{n-1} (x_{k} - y_{k})^{\ast}(x_{k} - y_{k})}\\
&=\sqrt{\left\langle x-y \middle| x-y \right\rangle}
\end{aligned}
\end{equation*}

We can also define the angle between two vectors, using the formula 
\begin{equation*}
\begin{aligned}
&\left\langle x \middle| y \right\rangle = \Vert \ket{x} \Vert \Vert \ket{y} \Vert \cos{\theta}\\
&\Longrightarrow \theta = \arccos{\left( \frac{\left\langle x \middle| y \right\rangle}{\Vert \ket{x} \Vert \Vert \ket{y} \Vert} \right)}
\end{aligned}
\end{equation*}
Clear definition of the distance and angle between two vectors allows generalization of linear algebra and calculus within a quantum state space.

The projective part comes from the nature of quantum mechanical systems where the coefficients $c_k \in \mathbb{C}$ defining a quantum state have to meet the condition $\sum \Vert c_k \Vert ^ 2 = 1$ \cite{qi_10}. For any quantum state that does not meet the condition, we can simply normalize it with the normalizing constant $\frac{1}{\sum \Vert c_k \Vert ^ 2}$. For a quantum state expressed as a unit vector $\ket{\psi}$, it is equivalent to any other state of the form $\gamma \ket{\psi}$ for some non-zero $\gamma \in \mathbb{C}$ i.e., $\frac{1}{\gamma}$ is the normalizing constant for $\gamma \ket{\psi} \in \mathbb{C}^n$, where $n=2$ for the qubit model and $n=\infty$ for the CV model. Here, $\gamma \in \mathbb{C}^{\ast}$ is called global phase and the original complex Hilbert space is modded out by $\mathbb{C}^{\ast}$, producing a complex "projective" Hilbert space.

\subsection{Discrete Variable Model}
The discrete variable model can be viewed as a quantized version of classical bit computing. 

\subsubsection{Quantum bits}
The classical bit states $0$ and $1$ are quantized into column vectors of length $2$ in a complex projective Hilbert space as computational basis $\{ \ket{0}, \ket{1} \}$ using Dirac notation. Any quantum state $\ket{\psi}$ within the computational space is called quantum bit (qubit) and expressed as a complex linear combination of $\ket{0}$ and $\ket{1}$ \cite{qi_10}. 
\begin{equation*}
  \ket{\psi} = \alpha \ket{0} + \beta \ket{1} 
 = \alpha \begin{bmatrix}
       1 \\ 0 \end{bmatrix}
  + \beta \begin{bmatrix}
       0 \\ 1 \end{bmatrix}
  = \begin{bmatrix}
       \alpha \\ \beta \end{bmatrix}
\end{equation*}

where $\alpha, \beta \in \mathbb{C}$ and $\vert \alpha \vert ^2 + \vert \beta \vert ^2 =1$. We have seen this normalized state is indeed a representative of the entire space $\mathbb{C}^{\ast}$ of global phases, giving us a projective Hilbert space of the spherical form, called the Bloch sphere. 

\begin{figure}[H]
    \centering
    \includegraphics[scale=0.5]{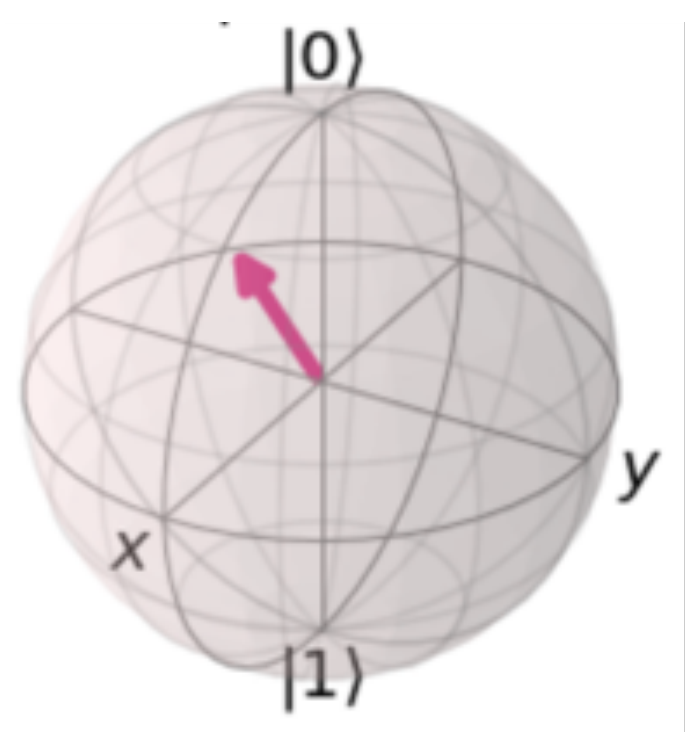}
    \caption{Quantum state space of a qubit}
    \label{fig:Bloch Sphere}
\end{figure}

The mathematical description of the Bloch sphere, which is a projective space of $\mathbb{C}^2-\{(0,0)\}$ modded out by the space of global phases, is given by $\mathbb{C}^{\ast}$
\begin{equation*}
\sfrac{\left(\mathbb{C}^2-\{(0,0)\}\right)}{\mathbb{C}^{\ast}} \cong
\sfrac{\left(S^3 \times \mathbb{R}^{+}\right)}{\left(U(1) \times \mathbb{R}^{+}\right)}
\cong \sfrac{S^3 }{U(1)} = \mathbb{C}P^{1}.
\end{equation*} 
$\mathbb{C}P^{1}$ represents the surface of the Bloch sphere.

\subsubsection{Multi qubits}
Increasing the computing power is achieved by adding more qubits. The quantum state in a system of multiple qubits is represented by the tensor product of the individual qubits. Let $\ket{\psi_k} = \alpha_k \ket{0} + \beta_k \ket{1}, k \in \{0,1,\ldots, n-1 \}$ be the quantum state of the $k^{th}$ qubit. Then any quantum state in the whole system is expressed as 

\begin{equation*}
\begin{aligned}
\ket{\psi_0} \otimes \ket{\psi_1} \otimes \ldots \ket{\psi_{n-1}}
&= \begin{bmatrix} \alpha_0\\ \beta_0 \end{bmatrix}
\otimes \begin{bmatrix} \alpha_1\\ \beta_1 \end{bmatrix}
\ldots 
\otimes \begin{bmatrix} \alpha_{n-1}\\ \beta_{n-1} \end{bmatrix}\\
&= \begin{bmatrix} \alpha_0\begin{bmatrix} \alpha_1 \ldots \begin{bmatrix} \alpha_{n-1}\\ \beta_{n-1} \end{bmatrix}\\ \beta_1 \ldots \begin{bmatrix} \alpha_{n-1}\\ \beta_{n-1} \end{bmatrix} \end{bmatrix} \\ 
\beta_0 \begin{bmatrix} \alpha_1 \ldots \begin{bmatrix} \alpha_{n-1}\\ \beta_{n-1} \end{bmatrix}\\ \beta_1 \ldots \begin{bmatrix} \alpha_{n-1}\\ \beta_{n-1} \end{bmatrix}\end{bmatrix}\end{bmatrix} \\
&= \begin{bmatrix}\alpha_0 \alpha_1 \ldots \alpha_{n-1}\\ 
\alpha_0 \alpha_1 \ldots \beta_{n-1}\\
\vdots \\
\beta_0 \beta_1 \ldots \alpha_{n-1}\\ 
\beta_0 \beta_1 \ldots \beta_{n-1} \end{bmatrix}.
\end{aligned}
\end{equation*} 

Each $\ket{\psi_k}$ is a vector of lenth $2$, hence the tensor product of $n$ vectors of length $2$ gives us a vector of length $2^n$. The resulting tensor product space is
\begin{equation*}
\mathbb{C}P^{n} = \sfrac{\left(\mathbb{C}^{n+1}-\{(0,0)\}\right)}{\mathbb{C}^{\ast}} 
\cong
\sfrac{\left(S^{2n+1} \times \mathbb{R}^{+}\right)}{\left(U(1) \times \mathbb{R}^{+}\right)}
\cong \sfrac{S^{2n+1}}{U(1)}.
\end{equation*} 

\subsubsection{Quantum gates}
Quantum gates induce change of states, representing information processing.
They are represented by $2 \times 2$ length-preserving rotation matrices on the Bloch sphere. They are of the form $UU^{\dagger}=U^{\dagger}U=I$ called unitary matrices, which means $U^{\dagger}= U^{-1}$. The inverse operation of a unitary matrix is the reverse rotation on the Bloch sphere. All quantum gates in the qubit model are reversible unlike some classical gates such as the AND and OR gates. 

The Hadamard gate $H$ is used to put the zero state $\ket{0}$ in uniform superposition of $\ket{0}$ and $\ket{1}$:

\[
H\ket{0} = \frac{1}{\sqrt{2}}
\begin{bmatrix}
1 & 1 \\ 1 & -1
\end{bmatrix}
\begin{bmatrix}
1 \\ 0
\end{bmatrix}
= \frac{1}{\sqrt{2}}
\begin{bmatrix}
1 \\ 1
\end{bmatrix}
= \frac{\ket{0} + \ket{1}}{\sqrt{2}}
\]

The Pauli-$X, Y, Z$ gates are used to induce half rotations about the $x, y$ and $z-$axis respectively:
\[
X (NOT) = 
\begin{bmatrix}
0 & 1 \\ 1 & 0
\end{bmatrix},
Y = 
\begin{bmatrix}
0 & -i \\ i & 0
\end{bmatrix},
Z = 
\begin{bmatrix}
1 & 0 \\ 0 & -1
\end{bmatrix}
\]
Parameterized gates are used to rotate a qubit state according to given parameters. Examples of parameterized gates are:
\[
RX(\theta) = 
\begin{bmatrix}
\cos \left(\frac{\theta}{2} \right) & -\sin \left(\frac{\theta}{2} \right) \\ -\sin \left(\frac{\theta}{2} \right) & \cos \left(\frac{\theta}{2} \right)
\end{bmatrix},
P(\theta) = 
\begin{bmatrix}
1 & 0 \\ 0 & e^{i \theta}
\end{bmatrix}
\]

The unitary gates acting on the entire system are represented by $2^n \times 2^n$ matrices as tensor products of $n$ $2 \times 2$ unitary matrices. For example, in a two qubit system when we apply the Hadamard gate on the first and the T gate on the second, the tensor product of the two matrices act on the entire system:

\begin{equation*}
\begin{aligned}
    H \otimes T 
    &= \frac{1}{\sqrt{2}}
    \begin{bmatrix} 1 & 1\\1 & -1 \end{bmatrix} \otimes
    \begin{bmatrix} 1 & 0\\0 & e^{i \frac{\pi}{4}} \end{bmatrix}\\
    &= \frac{1}{\sqrt{2}}\begin{bmatrix} 1 
    \begin{bmatrix} 1 & 0\\0 & e^{i \frac{\pi}{4}} \end{bmatrix}& 1\begin{bmatrix} 1 & 0\\0 & e^{i \frac{\pi}{4}} \end{bmatrix}\\
    1\begin{bmatrix} 1 & 0\\0 & e^{i \frac{\pi}{4}} \end{bmatrix} & -1\begin{bmatrix} 1 & 0\\0 & e^{i \frac{\pi}{4}} \end{bmatrix} \end{bmatrix}\\
    &= \frac{1}{\sqrt{2}}
    \begin{bmatrix} 1 & 0 & 1 & 0\\
    0 & e^{i \frac{\pi}{4}} & 0 & e^{i \frac{\pi}{4}}\\
    1 & 0 & -1 & 0 \\
    0 & e^{i \frac{\pi}{4}} & 0 & -e^{i \frac{\pi}{4}} \end{bmatrix}
\end{aligned}
\end{equation*}
acting on vectors of size $4$, representing 2-qubit states $\alpha_0 \ket{00} + \alpha_1 \ket{01} + \alpha_2 \ket{10} + \alpha_3 \ket{11}$, where $\sum \Vert \alpha_k \Vert^2 =1$ in the projective Hilbert space $\mathbb{C}P^{2}$.
In a multi-qubit system, we can use one or more qubits as a control for an operation on other qubits. For example, the Pauli$-X$ gate which is equivalent to the classical NOT gate on one qubit can be controlled by another qubit, as seen here.

\begin{figure}[H]
    \centering
    \includegraphics[scale=0.5]{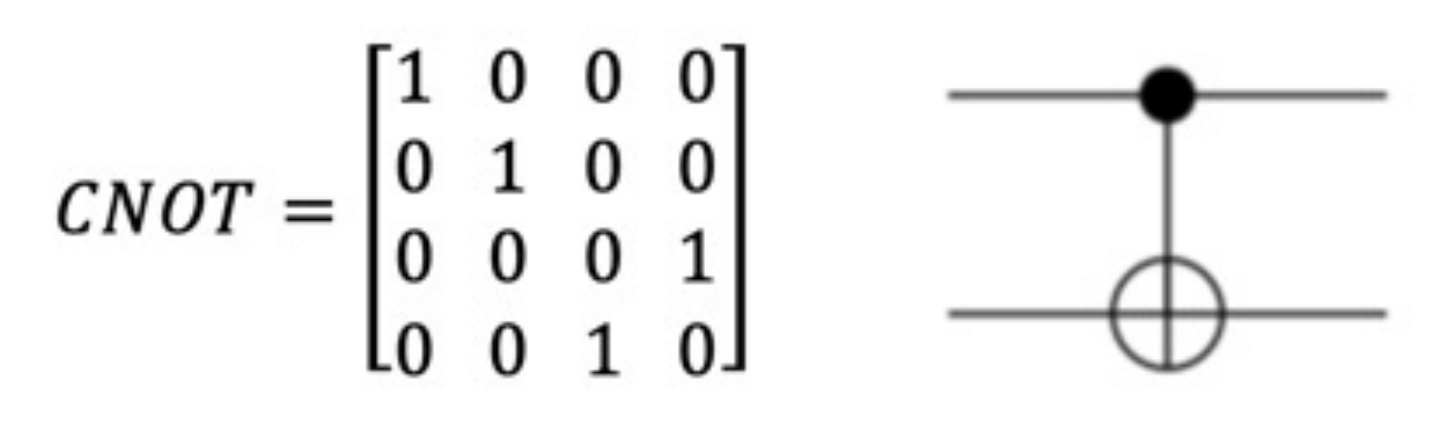}
    \caption{CNOT gate, i.e., controlled Pauli-X gate }
    \label{fig:CNOT}
\end{figure}

\subsubsection{Phase kickback}
Unlike in classical computing, the quantum state of the control qubit is affected by the state of the other qubit, on which the controlled operation is performed on. This phenomenon called phase kickback, unique to quantum computing, is used to record the evolution of the state change of a quantum circuit on the ancillary qubit. Phase kickback plays an important role in many quantum algorithms including quantum machine learning.

Let $U$ be an arbitrary unitary matrix and $\ket{\psi_k}$ be one of its eigenvectors. Then for its corresponding complex eigenvalue $\lambda_k = e^{i \alpha}$ expressed in Euler formula, $U$ acting on $\ket{\psi_k}$ is equal to multiplying the eigenvector by its eigenvalue $e^{i \alpha}: U\ket{\psi_k} = \lambda_k \ket{\psi_k} = e^{i \alpha} \ket{\psi_k} = e^{i \alpha} I \ket{\psi_k}$. Hence the effect of the gate operation $U$ on one of its eigenvectors is equivalent to the identity operation multiplied by its corresponding eigenvector. However, by the projective nature of the quantum state space, the eigenvalue acts as a global phase, returning us the original state as though no operation was applied. That picture changes when we apply a controlled $U$ gate instead. Consider a two-qubit circuit where an arbitrary unitary gate $U$ is applied to the second qubit controlled by the first. The $U$ operation will be applied only when the control qubit is in the state $\ket{1}$. The Hadamard gate is applied to the the first qubit, initialized to the $\ket{0}$ state, to obtain the uniform superposition of $\ket{0}$ and $\ket{1}$:

The states $\ket{0}$ and $\ket{1}$ can be expressed by their outer product matrices up to phase $\frac{1}{\sqrt{2}}$. The operation of the $U$ gate on the second qubit, when its state is one of the eigenvectors of $U$, conditioned on the $\ket{1}$ state of the first qubit is then given by

\begin{equation*}
\begin{aligned}
    &\ket{0} \bra{0} \otimes I + \ket{1} \bra{1} \otimes U
    = \ket{0} \bra{0} \otimes I + \ket{1} \bra{1} \otimes e^{i \alpha} I\\
    &= 
    \begin{bmatrix} 1 \\ 0 \end{bmatrix} 
    \begin{bmatrix} 1 & 0 \end{bmatrix} 
    \otimes 
    \begin{bmatrix} 1 & 0 \\0 & 1 \end{bmatrix} 
    + \begin{bmatrix} 0 \\ 1 \end{bmatrix} 
    \begin{bmatrix} 0 & 1 \end{bmatrix} 
    \otimes 
    \begin{bmatrix} e^{i \alpha} & 0 \\0 & e^{i \alpha} \end{bmatrix}  \\
     &= \begin{bmatrix} 1 & 0 & 0 & 0\\ 0 & 1 & 0 & 0 \\
     0 & 0 & 0 & 0\\ 0 & 0 & 0 & 0\end{bmatrix}
     + \begin{bmatrix} 0 & 0 & 0 & 0\\ 0 & 0 & 0 & 0 \\
     0 & 0 & e^{i \alpha} & 0\\ 0 & 0 & 0 & e^{i \alpha}\end{bmatrix} \\
     &=\begin{bmatrix} 1 & 0 & 0 & 0\\ 0 & 1 & 0 & 0 \\
     0 & 0 & e^{i \alpha} & 0\\ 0 & 0 & 0 & e^{i \alpha}\end{bmatrix}\\
    &=\begin{bmatrix} 1 & 0 \\ 0 &  e^{i \alpha} \end{bmatrix} 
     \otimes
     \begin{bmatrix} 1 & 0 \\ 0 & 1 \end{bmatrix}
\end{aligned}
\end{equation*}
Notice the first matrix which applies $e^{i \alpha}$ rotation on $\ket{1}$ on the first qubit and the identity matrix on the second. It is as though  the operation of $U=e^{i \alpha}I$ on the second qubit is kicked back to the first qubit when its state is $\ket{1}$.

When this operation is applied to $\frac{\ket{0} + \ket{1}}{\sqrt{2}} \otimes \ket{\psi}$, we get 
\begin{equation*}
    \begin{bmatrix} 1 & 0 \\ 0 &  e^{i \alpha} \end{bmatrix}
    \frac{1}{\sqrt{2}} \begin{bmatrix} 1 \\ 1 \end{bmatrix} \otimes \begin{bmatrix} 1 & 0 \\ 0 & 1 \end{bmatrix}\ket{\psi}
    = \frac{\ket{0} + e^{i \alpha}\ket{1}}{\sqrt{2}} \otimes \ket{\psi},
\end{equation*}
phase kick back from the second qubit to the first.

\subsubsection{Measurement}
The last stage of a quantum circuit is measurement, a read-out of computational results classically. It is a projection of the quantum computational resulting state onto one of the classical computational basis. Let 
\begin{equation*}
    \ket{\psi} = c_0\ket{00\ldots 0}+c_1\ket{00\ldots 1}+\ldots + c_{2^n-1} \ket{11\ldots 1}
\end{equation*} be the final state of computation in an $n-$qubit system. The probability of getting the $k^{th}$ computational basis state is given by “Born Rule” \cite{qi_10} as 
\begin{equation*}
  prob(k)= \frac{\vert c_k \vert^2}{\sum_j \vert c_j \vert^2} 
= \left\langle \phi \middle| k \right\rangle \left\langle k \middle| \phi \right\rangle
=\vert \left\langle k \middle| \phi \right\rangle \vert^2  
\end{equation*}

In each instance of measurement, we get one of the computational basis elements once. To extract the probability value of the $k^{th}$ element, which is a projection of $\ket{\psi}$ onto $\ket{k}$, we perform multiple shots of the circuit operation. 
\begin{figure}[b]
    \centering
    \includegraphics[scale=0.85]{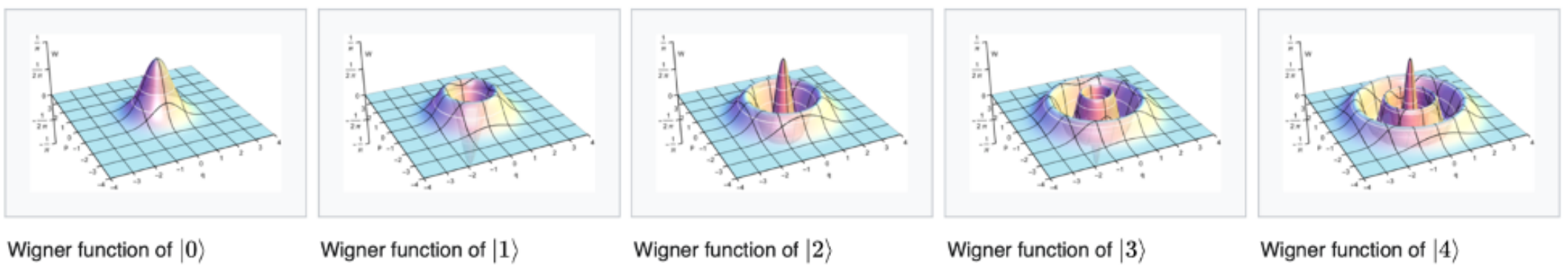}
    \caption{Fock basis  [\href{https://upload.wikimedia.org/wikipedia/commons/7/71/Wignerfunction_fock_0.png}{image source}]}
    \label{fig:Fock}
\end{figure}
\subsection{Continuous Variable}
The CV model is based on the wave-like property of nature and uses the quantum state of the electromagnetic field of bosonic modes (qumodes) as information carriers \cite{qi_cv_03}. Its physical implementation is done using linear optics, containing multiple qumodes \cite{optics_01}. In linear optics, the mathematical description of the electromagnetic field, based on different numbers of photons present inside qumodes, are used to represent the quantum state of of a system. 

A photon is a form of electromagnetic radiation, whose properties are described quantum mechanically. Phase space representation of the quantum state of a linear optics system describes the state using the position and momentum variables of the photons in the system. Depending on the wave function of the position variable and the momentum variable, we can derive a Gaussian distribution of the state on the position - momentum plane. The Wigner function is used which is a function of the position and momentum observables. It is given by
\begin{equation*}
    W(x,p)=\frac{~{p}}{h}=\frac{1}{h} \int_{-\infty}^{\infty}e^{-\frac{ipy}{\hbar}}\Psi \left(x+\frac{y}{2}\right)\Psi^{\ast}\left(x-\frac{y}{2}\right) dy
\end{equation*}
where $h=6.62607015 \times 10^{-34}$ is the Plank constant and $\hbar = 6.582119569 \times 10^{-16}$ the reduced Plank constant \cite{kitten_08}. The position wave function $\Psi(x)$ for each number of photons and its momentum $p$ is used on the position-momentum plane. Fock basis, i.e., the set of the images of the Wigner functions per the number of photons is depicted in Figure 3. Fock basis is derived from the position wave function and momentum, based on the number of photons present in the qumode. The position wave function describes the light wave electromagnetic strength of the qumode depending on the number of photons present. It is a complex valued function on real valued variables: $\Psi: \mathbb{R} \rightarrow \mathbb{C}: x \mapsto \alpha$. As the light waves from different photons interact either constructively or destructively, the wave function with more than one photon displays the constructive and destructive interactions between the light waves of the photons. 

\subsubsection{Qumode states}

The quantum state $\ket{\psi}$ of a qumode is expressed as a superposition of Fock basis states: 
\begin{equation*}
 \ket{\psi}=c_0 \ket{0}+c_1 \ket{1}+\hdots+c_n \ket{n}+\hdots \text{where} \sum_{k=0}^{\infty} \Vert c_k \Vert ^2=1
\end{equation*}
where $c_k$ is the probability amplitude of basis $\ket{k}$. 

In linear optics, this is physically obtained through the quanitization process of classical light called squeezing. A qumode is initialized to the vacuum state, i.e., zero photon state. The Gaussian distribution of the position and momentum observables of the zero photon state is clearly defined by the Wigner function. The uncertainty principle states the product of the standard deviation $\sigma_x$ of position $x$ and the standard deviation $\sigma_p$ of momentum $p$ has a clearly defined lower bound: $\sigma_x \sigma_p \ge \frac{\hbar}{2}$ for the reduced Plank's constant $\hbar = \frac{h}{2 \pi}$. Squeezing the momentum standard deviation $\sigma_p$ as close to zero as possible will increase the value of the position standard deviation because of the $\frac{\hbar}{2}$ lower bound.  The visual representation on the position-momentum plane is depicted in the figure.
\newpage
\begin{figure}[H]
    \centering
    \includegraphics[scale=0.5]{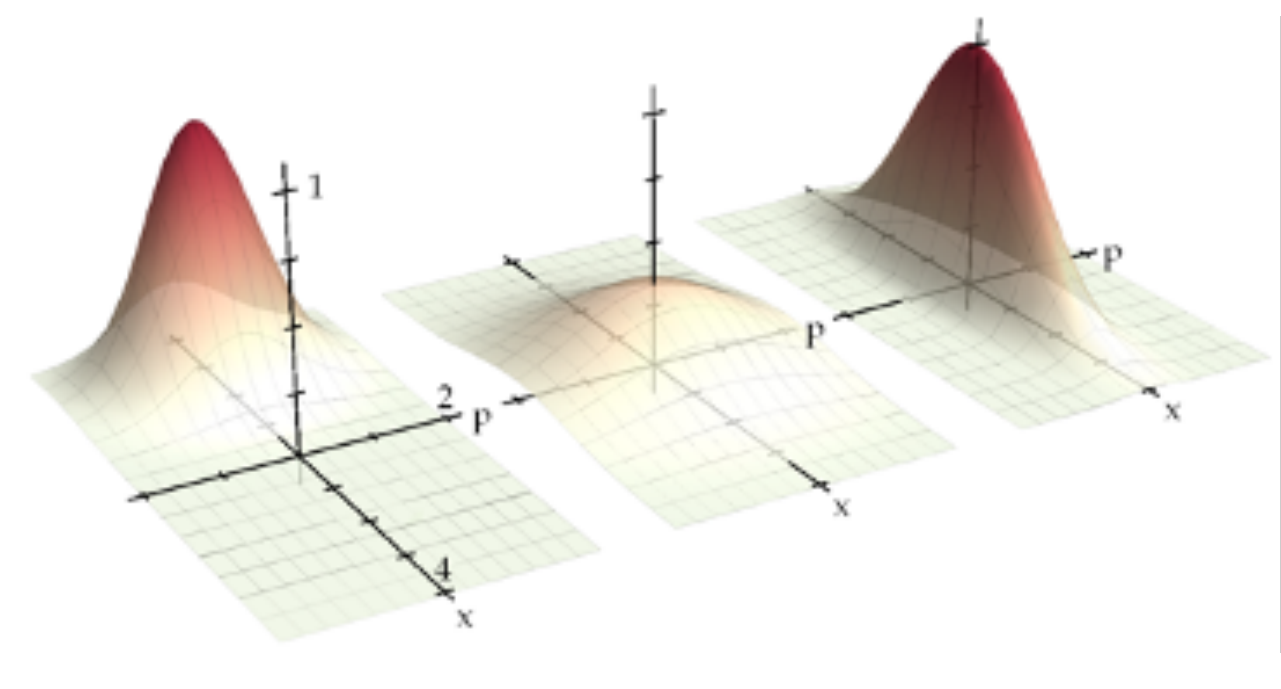}
    \caption{Squeezing: quantizing classical light \cite{kitten_08}}
    \label{fig:squeezing}
\end{figure}

As the value $\sigma_x$ spreads farther on the $x-$axis, the mathematical description of the qumode can be expressed as a superposition of the numbers of photons that can be found in the system, using Fock basis i.e., number basis. The probability of finding the qumode with $k-$photons is given by $\Vert c_k \Vert ^2$ when $\ket{\psi}$ is normalized. 

Available quantum information channels (qumodes) are initialized with no photons present, which are called vacuum states $\ket{0}$. The Wigner function $W_0(x,p)$ of each qumode is a Gaussian distribution depicting the lowest energy state. Laser beam light is applied to the qumodes and the classical light is converted to quantum light through a process called squeezing. It puts the deterministic classical light in superposition of Fock basis by squeezing the momentum observable of the light wave, thus spreading out the position observable. Application of a squeezer parameterized with the squeezing parameter $z$ on the vacuum state $\ket{0}$ is given by \cite{qi_12}
\[
S(z)\ket{0} = \frac{1}{\sqrt{\cosh{z}}} \sum_{n=0}^{\infty}\frac{\sqrt{(2n)!}}{2^n n!}\tanh^n{z} \ket{2n}.
\]
Notice the squeezing operation converted classical light into an infinite sum of Fock basis, i.e., quantum light.
\[
\ket{0} \rightarrow \sqrt{\frac{2}{e^z+e^{-z}}} \left(c_0 \ket{0} + c_2 \ket{2} + \ldots c_{2n} \ket{2n} + \ldots \right)
\]
where $c_{2k} = \frac{\sqrt{(2k)!}}{2^k k!}\left(\frac{e^z-e^{-z}}{e^z+e^{-z}}\right)^k$ for $k=0, 1, 2, \ldots$. The process can be visualized in the figure below.

A multi-qumode system is represented by the tensor product of individual qumode states: $\ket{\psi_0} \otimes \ket{\psi_1} \otimes \ldots \otimes \ket{\psi_{m-1}}$, where $m$ is the number of qumodes. When these states are approximated by cutoff dimension $n$, the computational basis of the resulting quantum state is of size $n^m$. 
\begin{equation*}
\begin{aligned}
   & \ket{\psi_0} \otimes \ket{\psi_1} \otimes \ldots \otimes \ket{\psi_{m-1}}=\\ 
   &d_0 \ket{00 \ldots 0} + d_1 \ket{00 \ldots 1} + c_{n^m-1}\ket{n-1,n-1,\ldots,n-1} 
\end{aligned}
\end{equation*}

\subsubsection{cutoff dimension}
As observed in the squeezed light diagram, the quantum state of a qumode expressed in Fock basis is centered around $0$, meaning it is very likely that $c_0 > c_1 > \ldots > c_n > \ldots$. Then we can approximate $\ket{\psi}$ with $\ket{\hat{\psi}}$ by cutting off the trailing terms. The number of Fock basis we use to approximate the true state $\ket{\psi}$ is called "cutoff dimension". Let $n$ be cutoff dimension. Then the approximating state $\ket{\hat{\psi}}$ is in superposition of $\ket{0}, \ket{1}, \ldots, \ket{n-1}$.
\begin{equation*}
 \ket{\psi}=c_0 \ket{0}+c_1 \ket{1}+\hdots+c_{n-1} \ket{n-1} \text{where} \sum_{k=0}^{n-1} \Vert c_k \Vert ^2=1
\end{equation*}

In vector representation, the state is expressed as a vector of size $n=$cutoff dimension.

\subsubsection{Quantum gates}
Fock basis states constitute the eigenstates of the number operator $\hat{n} := \hat{a}^{\dagger}\hat{a}$ where $\hat{a}^{\dagger}$ is called the constructor and $\hat{a}$ is called the annihilator \cite{qi_12}. Their matrix representation is given by

\begin{equation*}
\hat{a}^{\dagger} = 
\begin{bmatrix} 
0 & 0 & 0 & \hdots & 0 & 0\\
\sqrt{1} & 0 & 0 & \ldots & 0 & 0\\
0 & \sqrt{2} & 0 & \ldots & 0 & 0\\
0 & 0 & \sqrt{3} & \ldots & 0 & 0\\
\vdots &  & \ldots & \ddots &  & \vdots\\
0 & 0 & 0 & \ldots & \sqrt{n-1}& 0\\
\end{bmatrix}
\text{and}
\end{equation*}
\begin{equation*}
\hat{a} = 
\begin{bmatrix} 
0 & \sqrt{1} & 0 &0 & \ldots &  0\\
0 & 0 & \sqrt{2}& 0 & \ldots  & 0\\
0 & 0 & 0& \sqrt{3} & \ldots  & 0\\
\vdots &  & \ldots & \ddots &  & \vdots\\
0 & 0 & 0 & 0 &\ldots & \sqrt{n-1}\\
0 & 0 & 0 & 0 &\ldots &  0\\
\end{bmatrix}
\end{equation*}

Note that
\begin{gather*}
\hat{a} \ket{0} = 0, \hat{a} \ket{k} = \sqrt{k} \ket{k-1} \text{ for }k \ge 1 \text{ and}\\
\hat{a}^{\dagger} \ket{k} = \sqrt{k+1} \ket{k+1} \text{ for }k \ge 0.
\end{gather*}

The product of $\hat{a}^{\dagger}$ and  $\hat{a}$ returns a matrix 
\begin{equation*}
   \begin{bmatrix} 
0 & 0 & 0 &0 & \ldots &  0\\
0 & 1 & 0 & 0 & \ldots  & 0\\
0 & 0 & 2& 0 & \ldots  & 0\\
\vdots &  & \ldots & \ddots &  & \vdots\\
0 & 0 & 0 & 0 & n-2 & 0\\
0 & 0 & 0 & 0 & 0 & n-1
\end{bmatrix} 
\end{equation*}
which we denote "number operator" $\hat{n}$. Then $\hat{n}\ket{k} = k \ket{k}$ where $k \in \{ 0, 1, \ldots, n-1 \}$ for cutoff dimension$=n$.  

Standard Gaussian gates in the CV model are of the form $U= e^{-iHt}$ where $H$ represents the Hamiltonian of the system, describing the total energy as the sum of kinetic and potential energy. The matrix exponential $e^{-iHt}$ is of the form
\begin{equation*}
e^{-iHt}= \sum_{k=0}^{\infty}\frac{({-iHy})^k}{k!}=I-iHt + \frac{(-iHt)^2}{2!}+ \hdots + \frac{{-iHt}^n}{n!}+ \hdots
\end{equation*}
In the finite quantum state space of a qumode approximated by cutoff dimension $n$, this infinite sum representing a quantum gate is also approximated by a finite matrix exponential.

\begin{equation*}
\hat{U} = \sum_{k=0}^{n-1}\frac{({-iHt})^k}{k!}=I -iHt + \frac{(-iHt)^2}{2!}+ \hdots + \frac{{-iHt}^{n-1}}{(n-1)!}
\end{equation*}

Some of the standard Gaussian gates taking Gaussian states to Gaussian states are listed below.

Squeezer with parameter $z: S(z) = exp \left( \frac{z^{\ast} \hat{a}^2+z \hat{a}^{{\dagger}^2}}{2} \right)$

Rotation with parameter $\phi: R(\phi) = exp\left(i \phi \hat{a}^{\dagger}\hat{a} \right)$

Displacement with parameter $\alpha: D(\alpha)=exp\left(\alpha \hat{a}^{\dagger}-\alpha^{\ast} \hat{a} \right)$

Another Gaussian gate, the beamsplitter, is a two-qumode gate with parameters $\theta$ and $\phi$: 
\[
B(\theta, \phi) =
exp\left( \theta \left( e^{i\phi} \hat{a} \hat{b}^{\dagger} + e^{-i \phi}
\hat{a}^{\dagger}\hat{b} \right) \right)
\]
where $\hat{b}^{\dagger}$ and $\hat{b}$ are constructor and annihilator of the 2nd qumode respectively.

On an $m-$qumode system, $m-1$ beamsplitters on pairs of adjacent qumodes and $m$ rotation matrices form an interferometer.

\begin{figure}[H]
    \centering
    \includegraphics[scale=0.5]{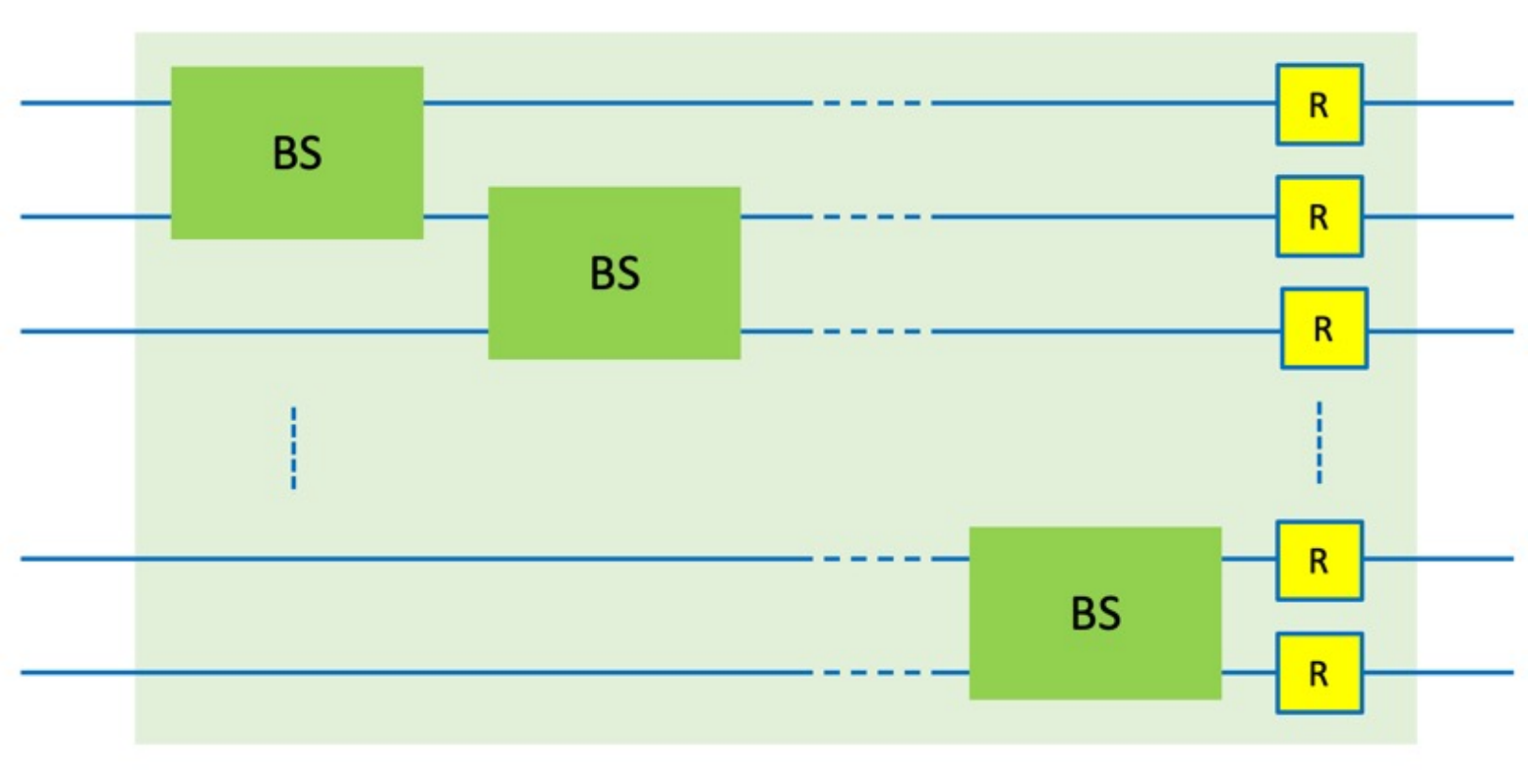}
    \caption{Components of an interferometer: beamsplitters and rotations matrices}
    \label{fig:qnn}
\end{figure}

\subsubsection{Measurement}
Measurement is done via counting the number of photons present in each qumode with a photon detector. Xanadu's Pennylane offers a suite of measurement methods as outlines in Table 1. The size of the output varies from 1 to the number of shots, the number of qumodes, and its square. Observe that varying the number of qumodes affects the size of the output vectors of a quantum circuit. In addition, the notion of cutoff dimension can be used to extract output vectors of desired size. Currently available QPU by Xanadu, $X8$, offers $8$ qumodes. There are several ways of getting output vectors of size $n \le 8$: one qumode with cutoff dimension $n$, $n$ qumodes, and $Int(log_{2} n)$. 

\begin{table}
\centering
\begin{tabular}[H]{lcc}
\hline
Measurement method & output size & output size \\
& per qumode & for $m-$qumodes \\ \hline
expectation value & $1$ & $m$\\ 
variance   & $1$ & $m$\\
probability & $n$ & $n^m$\\
\hline
\end{tabular}
\end{table}

Let $\ket{\psi} = \begin{bmatrix} \psi_0 \\ \psi_1 \end{bmatrix}$ be the quantum state of a multi-qumode system after desired quantum computational operations are performed. The expectation value measurement method returns $\bra{\psi} A \ket{\psi}$ where the operator $A$ is usually the Pauli$-X$, Pauli$-Y$, or Pauli$-Z$ gate. The expectation value of the Pauli$-X$ matrix is
\begin{equation*}
\begin{aligned}
    \bra{\psi} X \ket{\psi} &= 
    \begin{bmatrix} \psi_0^{\ast} & \psi_1^{\ast} \end{bmatrix}
    \begin{bmatrix} 0 & 1 \\ 1 & 0 \end{bmatrix}
    \begin{bmatrix} \psi_0 \\ \psi_1 \end{bmatrix} \\
    &= 
    \begin{bmatrix} \psi_0^{\ast} & \psi_1^{\ast} \end{bmatrix}
    \begin{bmatrix} \psi_1 \\ \psi_0 \end{bmatrix}\\
    &= 
    \psi_0^{\ast} \psi_1 + \psi_1^{\ast} \psi_0 \\
    &=
    2\left( Re(\psi_0)Re(\psi_1) + Im(\psi_0)Im(\psi_1) \right)\in \mathbb{R},
\end{aligned}
\end{equation*}
which is a real number. In a multi-qumode system with $m$ qumodes, we can get a vector of size $m$ by getting the expectation value for each qumode or one single value by getting the dot product of the expectation values, which simply is the product of all of them. 

With the variance method, we get a vector of size $m$ with each entry representing the measure of squared deviation from the sample mean.

The probabilities method returns the probability of each computation basis state. Suppose an $m-$qumode system has cutoff dimension $n$ for each qumode. Then, there are $n^m$ computational basis states, hence we get a vector of size $n^m$.

%% file: imp.tex
\section{TYPES OF QUANTUM COMPUTERS}

The models of physical quantum computers that are available can be divided into two groups: the adiabatic model implemented by D-wave and the gate model implemented by IBM, Google, Xanadu, and other companies. 

\subsection{Adiabatic quantum computing}
In adiabatic quantum computing, the overall energy of a physical system is used for information processing. D-wave made headlines with its adiabatic quantum computer in which the Hamiltonian of the given system is used to find the global minimum solution for an objective function. Adiabatic quantum computing uses a gradual process of evolving the energy of a quantum mechanical system from the initial state to the state describing the solution to a given problem. It is well suited for optimization and sampling problems. 
The total energy of a quantum mechanical system, both kinetic and potential, can be mathematically described by a function called Hamiltonian. It describes the energy of a system, as a function of the position and momentum of a particle, by mapping eigenstates to energies. Quantum annealing is the process of evolving the initial energy state to the global minimum solution state of an objective function. In physical systems, this ideal process is realized via an adiabatic process, which is a slow and gradual annealing process without interference from outside energy sources.

Manipulating the Hamiltonian of the overall system does not have readily direct correlations to the known classical algorithms. Therefore a lot of quantized versions of classical algorithms are better implemented in gate model quantum computing.

\subsection{Gate model quantum computing}

In gate model quantum computing, the quantum states of individual information carriers are used as units of information. The state is mathematically expressed as a vector, which is a complex linear combination of the quantized classical computational basis, in a complex projective Hilbert space. Controlled change of states of the information carriers is translated as a quantum logic gate, mathematically represented by a matrix \cite{qi_10}. Quantum gates, unlike classical logic gates, are reversible. A series of quantum gates is called a quantum circuit. The computational results of gate operations are accessed classically via an operation called measurement. 

From the physical implementation perspective, quantum gates are realized by controlling either electrical or photonic waveforms \cite{qc_19}. Physical implementation of the gate model can be done via superconducting based on the discrete variable model, linear optics (photonics) based on the CV model, and ion trap, among other means that are being pursued. Although Honeywell and IonQ are working hard on ion trap, their chips have not been made available to the public. IBM and Google have cloud services of their superconducting chips and Xanadu also has of their photonic chips. 

\subsubsection{Superconducting model}

In the superconducting model, the quantum information carrying agent is an individual particle \cite{qc_19}. Superconductor is a unique class of materials that exhibit no electrical resistance at zero frequency when cooled to below a critical temperature. From a collection of multiple quantized energy states of the superconductor, the two lowest energy states can be selectively accessed to realize the qubit: the ground state as the state $\ket{0}$ and the first excited state as $\ket{1}$ \cite{qc_19}. 

The quantum state of the particle is manipulated by microwave for computation \cite{QML_14}. A series of operations is performed on the particle for information processing. Precise control of its state for computation requires a supercondensing low temperature to avoid unwanted thermal excitation of the excited state. For Google’s quantum chip, for example, it is cooled below 20 mK, using a dilution refrigerator. This sensitivity of a quantum particle to its environment is called a decoherence problem. The decoherence problem is something that needs to be addressed in superconducting quantum computers on the way to fault-tolerant universal quantum computing. 

For read-out, each particle needs to be connected to classical computers via a wire. Hence as the number of computational particles is increased, the size of the accompanying hardware grows. The interaction between the classical information from room temperature and the quantum data plane connected via wires inside a dilution refrigerator may cause unwanted thermal noise \cite{qc_19}. 

Since the quantum state of the overall system is based on a cluster of individual particles, each of which state is represented by a qubit, this is a natural implementation of the qubit model of quantum computing.
	
\subsubsection{Linear optics model}	
Linear optics quantum computing model uses the electromagnetic field of bosonic modes as quantum information carriers. The quantum state of the overall system containing multiple qumodes is defined by the interference of photonic electromagnetic field of the qumodes. As the mathematical description of the electromagnetic field is continuous, linear optics naturally implements the CV model of quantum computing: optical instruments to carry out quantum computations and photon detectors to measure and store the computational results \cite{photon_20}. 

Quantum gates are realized by controlling the ambient electromagnetic field of the qumodes using optical instruments. After computation, the resulting quantum states are accessed via photon detectors, by counting the number of photons in each qumode. The optical instruments function at room-temperature and are not susceptible to decoherence problems. Without the need for a huge dilution refrigerator, it can work on any regular size computer, hence can be a regular stand-alone server. It is easily scalable, since it can be embedded into the current fiber optics infrastructure \cite{boson_21}. 

The table below outlines the comparison between the superconducting model and the linear optics model.\\

\begin{table}[htbp]
  \centering
  \caption{Comparison of Superconducting and Photonic Quantum Computers}
  \resizebox{0.7\textwidth}{!}{\begin{minipage}{\textwidth}

\begin{tabular}{ |c|c|c|}
\hline
    & Superconducting & Photonic \\ \hline
    properties &	particle like &	wave like\\ \hline
units of computation&	quantum bits&	quantum modes  \\ \hline
single mode & {$\ket{0}, \ket{1}$}& {$\ket{0}, \ket{1}, \ldots, \ket{n}, \ldots$} \\ 
computational basis& quantized digital binaries& Fock (number) basis   \\ \hline
basis size& 2& infinite   \\ \hline
multi mode & $\{\ket{0}, \ket{1}\}^{\otimes^m}$& $\{\ket{0}, \ket{1}, \ldots, \ket{n}, \ldots\}^{\otimes^m}$ \\ 
computational basis& tensor product & tensor product   \\ \hline
basis size& $2^m$& infinite   \\ \hline
state space& Hilbert space& Fock space   \\ 
& & Direct sum of Hilbert spaces   \\ \hline
physical implementation& electron & photons \\ \hline
temperature& below 20 mK & room temperature \\ \hline
server& dilution refrigerator & regular server \\ \hline
\end{tabular}
\end{minipage}}
\end{table}

%% file: QPU.tex
\section{QUANTUM PROCESSING UNITS}

Under the gate model of quantum computing, we examined the difference between the superconducting model and the linear optics model. IBM's Honeycomb and Google's Sycamore are superconducting QPUs and Xanadu's X8 is an 8-qumode linear optical QPU. IBM just unveiled a 127 qubit QPU called Eagle, allowing for up to a $2^127 \sim 10^38$-dimensional Hilbert space for quantum computation. Google’s Sycamore offers 63 qubits, allowing for up to a $2^63 \sim 10^19$-dimensional Hilbert space. Xanadu’s X8 has 8 qumodes, each of which can be set at a certain cutoff dimension. For building quantum circuits, Google offers a software package called Cirq, IBM Qiskit, and Xanadu Strawberry Fields. To better understand the mechanics of superconducting and photonic QPUs, we have a closer look at Google’s Sycamore and Xanadu’s X8.

\subsection{Sycamore} 

Google's $63-$qubit Sycamore, whose architecture is of a lattice shape as shown in the figure. To select $n-$qubits from the chip, it is customary to pick a $log_2 (n) \times log_2 (n)$ block.

Google’s 63 qubit QPU sycamore uses aluminum for metallization and a thin layer of non-superconducting indium for bump-bonds between two silicon wafers. Conducting electrons on the wafers are condensed to macroscopic quantum state, such that currents and voltage behave quantum mechanically. In order to achieve that, the chips are cooled to below 20 mK in a dilution refrigerator. 

The architecture of the chip is of lattice structure where each node represents a qubit. It is composed of nonlinear resonators at 5 - 7 Ghz. As controls, a microwave drive is used to excite the qubit and a magnetic flux control to tune the frequency. 

\begin{figure}[h]
    \centering
    \includegraphics[scale=0.5]{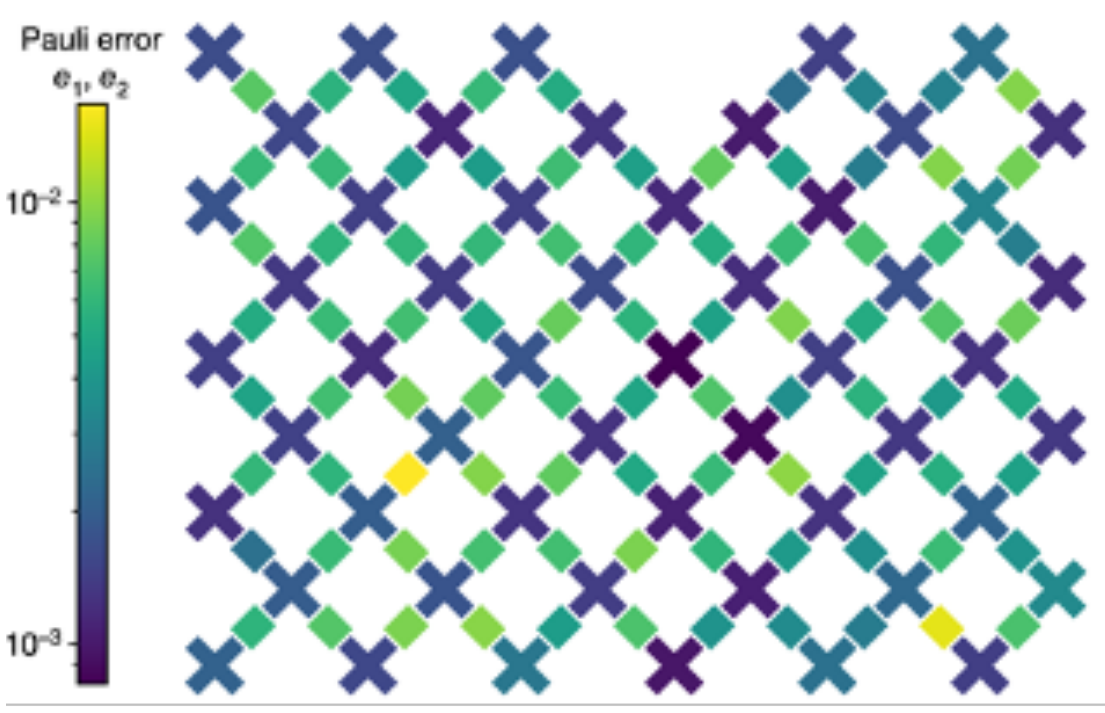}
    \caption{Lattice structure of Sycamore chip}
    \label{fig:Sycamore}
\end{figure}

For readout, a linear resonator connected to each qubit simultaneous readout using frequency- multiplexing technique. The software package offered for quantum circuits is Cirq, based on Python.

\subsection{X8} 

Xanadu’s 8-qumode photonic QPU is made of silicon nitrade for photon conductance. It is housed in a conventional server at room temperature and is accessible to the user via Python-based Strawberry Fields from a personal computer. A master controller, connected to a personal computer, controls the number of qumodes to be used, quantum circuit building, information processing, and read-out of the measurement results back to the user. For computation, classical light is pumped into the qumodes on the chip, converted into quantum light, goes through a series of quantum state changes based on a user-defined quantum circuit, filtered, and measured. The functional components of X8 are Pump I/O, pump distribution, squeezing, filtering, interferometer and programmable quantum gates. Pump I/O is a custom modulated pump laser source that produces a regular pulse train. The pump distributor directs the photons in classical states to appropriate qumodes.

\begin{figure}[h]
    \centering
    \includegraphics[scale=0.5]{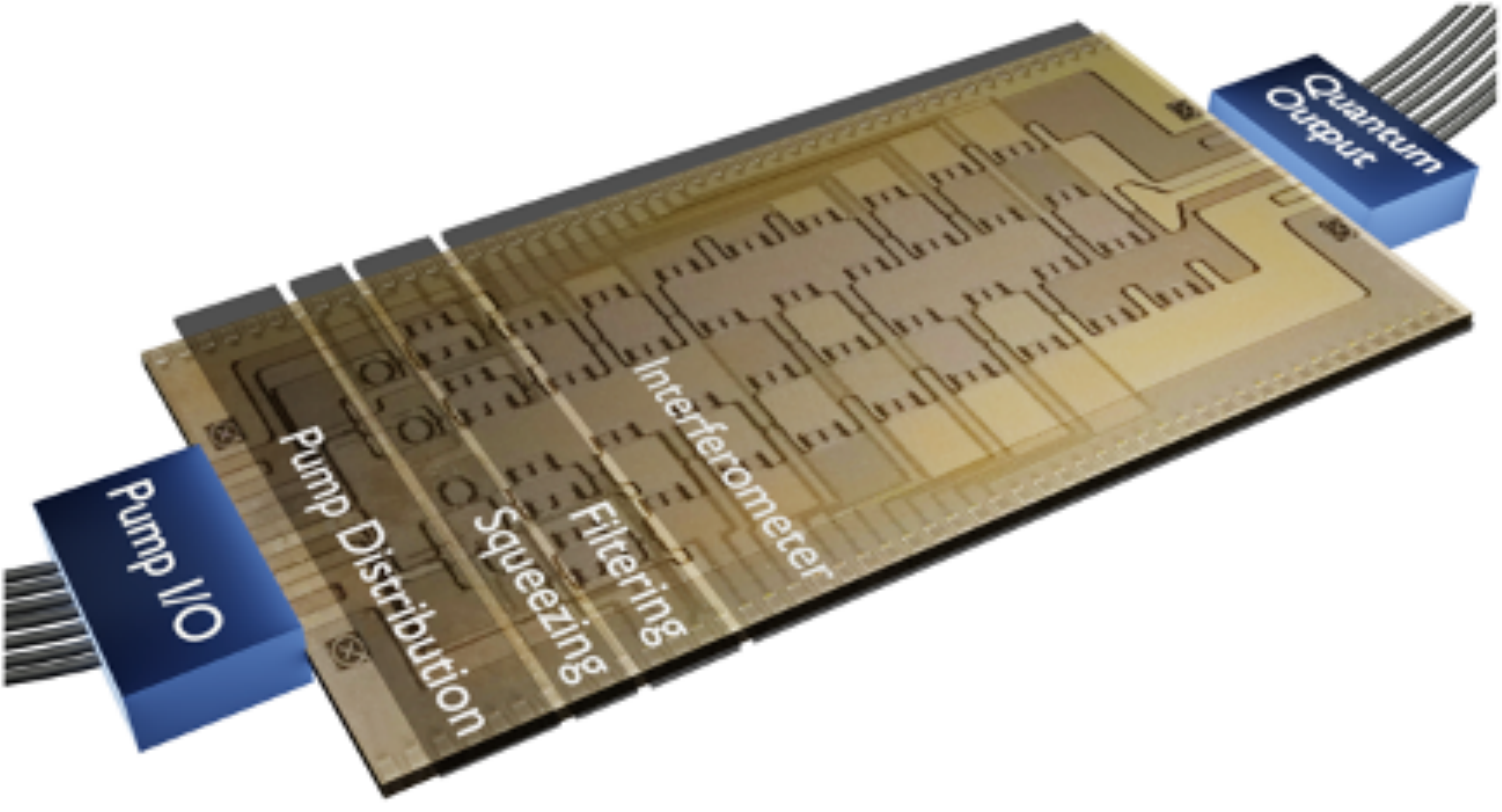}
    \caption{Xanadu's 8-qumode chip}
    \label{fig:X8}
\end{figure}

The process of squeezing converts the classical light into quantum squeezed state. 

The interferometer portion of the chip performs quantum circuit operations. The output from the chip is run through a filter to suppress unwanted light, passing only wavelengths close to the signal. Then the photon counter reads the number of filtered photons in each qumode and sends the result of the computation back to the user via master controller.

Photonic quantum computing has several advantages over superconducting quantum computing. The most important is that it can be operated at room temperature. It is also compatible with existing optical infra structure, hence networking and multiplexing would be a natural extension of the existing infrastructure. It is robust to decoherence, hence offers near perfect phase stability. The dimension of the entangled state grows exponentially with the number of photons and the number of modes.

%% file: conc.tex
\section{CONCLUSION}
We are at the beginning stage of a new computing paradigm where special purpose quantum computing will solve a class of problems that were not easily solvable in classical computing. The near term devices available on cloud are enabling researchers to test and develop new quantum algorithms in the areas of quantum chemistry, Gaussian boson sampling, graph optimization, and quantum machine learning. 

The near-term devices are still limited in terms of width and depth of quantum circuits that can carry out stable quantum computations. Due to their limited number of qubits/ qumodes, there is a challenge in encoding practical classical data into quantum states. Although there are numerous quantum algorithms with successful experimental results, quantum advantage over classical is yet to be actualized. As QPUs with more qubits or qumodes are being developed, corresponding quantum algorithms with efficient data encoding schemes and circuits can be tested and evaluated.

This paper examined the difference between qubit-based superconducting quantum computers and CV-based photonic quantum computers. The qubit based model works with the quantum mechanical states of individual particles while the CV model works with the quantum mechanical states of the electromagnetic fields of qumodes containing photons. Apart from operating at room temperature with linear optics instruments, photonic quantum computers hold much promise because of the infinite demensionality and rich array of quantum gates the CV model offers, on which photonic quantum computers are theoretically based on.

\section*{ACKNOWLEDGMENTS}
I appreciate assistance and guidance from Maria Schuld and those at Xanadu.